\documentclass[%
 aip,
 amsmath,amssymb,
 reprint,%
onecolumn]{revtex4-2}

\usepackage{graphicx}% Include figure files
\usepackage{dcolumn}% Align table columns on decimal point
\usepackage{bm}% bold math
\usepackage[utf8]{inputenc}
\usepackage[T1]{fontenc}
\usepackage{mathptmx}
\usepackage{etoolbox}

\makeatletter
\def\@email#1#2{%
 \endgroup
 \patchcmd{\titleblock@produce}
  {\frontmatter@RRAPformat}
  {\frontmatter@RRAPformat{\produce@RRAP{*#1\href{mailto:#2}{#2}}}\frontmatter@RRAPformat}
  {}{}
}%
\makeatother
\begin{document}

\preprint{AIP/123-QED}

%\preprint{}

\title{Hydrodynamic modulation instability triggered by a two-wave system} %Title of paper


\author{Yuchen He}\email{yuchen.he@polyu.edu.hk} 
\affiliation{Department of Civil and Environmental Engineering, The Hong Kong Polytechnic University, Hong Kong SAR}%
\affiliation{Department of Ocean Science and Engineering, Southern University of Science and Technology, Shenzhen 518055, China}%
\author{Jinghua Wang}
\affiliation{Department of Civil and Environmental Engineering, The Hong Kong Polytechnic University, Hong Kong SAR}%
\affiliation{Research Institute for Sustainable Urban Development, The Hong Kong Polytechnic University, Hong Kong SAR}%
\affiliation{Shenzhen Research Institute, The Hong Kong Polytechnic University, Shenzhen 518057, China}%
\author{Bertrand Kibler} 
\affiliation{Laboratoire Interdisciplinaire Carnot de Bourgogne, UMR6303 CNRS-UBFC, 21000 Dijon, France}%
\author{Amin Chabchoub}
 \email{chabchoub.amin.8w@kyoto-u.ac.jp} 
 \affiliation{Hakubi Center for Advanced Research and Disaster Prevention Research Institute, Kyoto University, Kyoto, Kyoto 606-8501, Japan}
\affiliation{Department of Infrastructure Engineering, The University of Melbourne, Parkville, Victoria 3010, Australia}%
%\author{}
%\thanks{}

\date{\today}

\begin{abstract}
The modulation instability (MI) is responsible for the disintegration of a regular nonlinear wave train and can lead to strong localizations in a from of rogue waves. This mechanism has been studied in a variety of nonlinear dispersive media, such as hydrodynamics, optics, plasma, mechanical systems, electric transmission lines, and Bose-Einstein condensates, while its impact on applied sciences is steadily growing. Following the linear stability analysis of weakly nonlinear waves, the classical MI dynamics, can be triggered when a pair of small-amplitude sidebands are excited within a particular frequency range around the main peak frequency. That is, a three-wave system is usually required to initiate the wave focusing process. Breather solutions of the nonlinear Schrödinger equation (NLSE) revealed that MI can generate much more complex localized structures, beyond the three-wave system initialization approach or by means of a continuous spectrum. In this work, we report an experimental study for deep-water surface gravity waves asserting that a MI process can be triggered by a single unstable sideband only, and thus, from a two-wave process when including the contribution of the peak frequency. The experimental data are validated against fully nonlinear hydrodynamic numerical wave tank simulations and show very good agreement. The long-term evolution of such unstable wave trains shows a distinct shift in the recurrent Fermi-Pasta-Ulam-Tsingou focusing cycles, which are captured by the NLSE and fully nonlinear hydrodynamic simulations with minor distinctions.\end{abstract}

%\pacs{47.85.Dh}% insert suggested PACS numbers in braces on next line

\maketitle %\maketitle must follow title, authors, abstract and \pacs
\begin{quotation}
Localized wave patterns in nonlinear dispersive media can occur as a result of modulation instability, which is triggered by small perturbations of frequency and amplitude. The ensuing even wave transfer ensures the broadening of wave spectrum and focusing of the wave field in physical space. This is also known as the Benjamin-Feir instability or quasi four-wave resonant interactions in the context of surface gravity and ocean waves. Following the linear stability of Benjamin and Feir the initialization of the process requires a symmetric sideband perturbation of Stokes waves. We provide experimental evidence for water waves that this fundamental instability can also be precipitated by a single sideband seeding. Despite the formation of the opposite sidebands at a later stage, the long-term evolution of the unstable wave field reveals shifted focusing recurrence cycles. \end{quotation}
\section{Introduction}
Waves in nonlinear dispersive media can become unstable when subjected to long-wave perturbations \cite{benney1967propagation,remoissenet2013waves,osborne2002nonlinear,ablowitz2011nonlinear}. Such an intriguing transfer of energy process from the peak frequency to sidebands is known as modulation instability (MI) and is not unique to water waves, in which it was first experimentally observed \cite{benjamin1967instability,bailung1993observation,tai1986observation,everitt2017observation}. The result of the linear stability analysis of second-order Stokes waves provides an exact range of unstable frequencies in which a pair of small-amplitude sidebands around the peak frequency will start to grow exponentially resulting in a disintegration and periodic wave group focusing of the wave field \cite{zakharov2009modulation}. The same result can be obtained if applying the linear stability analysis to the regular envelope solution of the nonlinear Schrödinger equation (NLSE) \cite{yuen1982nonlinear,osborne2002nonlinear}. Despite its numerous physical limitations, investigating the MI using the NLSE at that order of approximation in wave steepness is more diligent since the framework can reasonably predict the saturation of sidebands' growth, their decay, and the follow-up Fermi-Pasta-Ulam-Tsingou (FPUT) focusing recurrence \cite{fermi1955studies,yuen1982nonlinear,trillo1991dynamics,tulin1999laboratory,chiang2005theory,dauxois2008fermi,babanin2011breaking,kimmoun2016modulation,pierangeli2018observation,mussot2018fibre}. Moreover, exact solutions, such as the famed Akhmediev (ABs) and Peregrine breathers \cite{akhmediev1985generation,peregrine1983water}, describe the nonlinear stage of modulation instability and are useful to trigger and control the dynamics of modulationally unstable waves in laboratory environments \cite{dudley2009modulation,kibler2010peregrine,chabchoub2011rogue,bailung2011observation,chabchoub2014hydrodynamics,luo2020creation,romero2024experimental}. Note that doubly periodic A-type ABs can also describe the MI outside the conventional MI instability band \cite{akhmediev1986modulation,conforti2020doubly,vanderhaegen2021extraordinary}. 

Here, we report results of an experimental campaign aiming at the investigation of the initialization of MI for surface gravity waves from a two-wave system, i.e., a carrier frequency and only one small-amplitude sideband inclusion instead of exciting a pair of two, as suggested from the linear stability analysis. To initiate the evolution either in our numerical or physical wave tank, we use both, a spectrally truncated AB formalism, which consists of removing all higher or lower  frequencies with the respect to the peak frequency, and a single-seeded sideband to a second-order Stokes wave. Similarly to previous experiments in optics \cite{tai1986generation,kibler2010peregrine}, we confirm that one sideband is sufficient to trigger the instability dynamics in hydrodynamics despite a slight delay in the focusing dynamics compared to the classical three-wave excitation. The experiments are in very good agreement with the fully nonlinear numerical wave tank simulations. We also investigate the long-term behavior of unstable wave envelopes and show that these undergo a phase-shifted FPUT focusing recurrence \cite{kimmoun2016modulation,alberello2023dynamics} compared to the conservation, classical, and non-shifted MI dynamics. 

\section{Formalism} 
In their pioneering work, Benjamin and Feir showed that a second-order Stokes wave of amplitude $a$, wave frequency $\omega$, and wavenumber $k$ 
\begin{eqnarray} 
\eta_S(x,t)=a \cos\left[kx-\omega t\right]+\dfrac{1}{2}ka^2\cos\left[2\left(kx-\omega t\right)\right]
\end{eqnarray}
is unstable to long-wave perturbation in the temporal domain, if the sidebands are triggered within the bounded modulation frequency range \cite{benjamin1967disintegration,zakharov2009modulation} 
\begin{eqnarray} 
0<\Omega<\sqrt{2}ka\omega.
\end{eqnarray}
Intriguingly, the same condition can be found when applying the linear stability analysis on the constant and steady background envelope Stokes solution $\psi_S(x,t)=a\exp(ia^2k^3 x)$ of the time-like NLSE \cite{yuen1982nonlinear,osborne2002nonlinear}
\begin{equation}
i(\frac{\partial \psi}{\partial x}+\frac{1}{c_g}\frac{\partial \psi}{\partial t}) - \frac{k}{\omega^2} \frac{\partial^2 \psi}{\partial t^2} - k^3 |\psi|^2 \psi = 0.
\label{NLSE}
\end{equation}
Taking advantage of the integrability of the NLSE (\ref{NLSE}) \cite{shabat1972exact,akhmediev1997solitons,dysthe1999note}, classes of exact solutions can describe the MI process beyond the non-physical predictions of the linear stability analysis, which for instance does not imply any saturation of the exponentially growing sidebands. In fact, the well-known ABs, parametrized as \cite{akhmediev1985generation,dudley2009modulation}
\begin{eqnarray} 
\psi_{AB}(x,t)=a\left(1+\dfrac{2\left(1-2\mathfrak{a}\right)\cosh\left(a^2k^3 \mathfrak{b}x\right)+i\mathfrak{b}\sinh\left(a^2k^3\mathfrak{b} x\right)}{\sqrt{2\mathfrak{a}}\cos\left(\dfrac{ak}{\sqrt{2}\omega}\Omega \left(t-\dfrac{x}{c_g}\right)\right)-\cosh\left(a^2k^3\mathfrak{b} x\right)}\right)\exp(ia^2k^3 x),
\label{breather}
\end{eqnarray} 
where $\mathfrak{b}=\sqrt{8\mathfrak{a}(1-2\mathfrak{a})}$ and $\mathfrak{a}$ is growth rate control parameter, accurately describe the complete MI focusing process, involving an infinite number of sidebands, and the FPUT recurrence \cite{akhmediev1985generation,wetzel2011new}. Over decades, numerical and laboratory experiments studying this classical universal instability have brought new insights and widened the application prospects to an {\it old} problem \cite{onorato2006modulational,waseda2009evolution,galchenko2010modulational,kharif2010under,chabchoub2014hydrodynamics,kimmoun2016modulation,steer2020experimental,li2011modulation,mendes2022saturation,li2023coupled,andrade2023nonlinear,li2024currents}. To ensure a rapid growth in the numerical and physical tank, we chose $\mathfrak{a}=0.25$, which corresponds to the case of maximal growth rate \cite{dudley2009modulation}. In hydrodynamics up to now, the initial or boundary conditions employed to initiate the instability dynamics were steadily symmetric, i.e., a pair of small-amplitude sidebands were employed to destabilize the regular or quasi-regular wave field. That said, experiments in optics \cite{tai1986generation,kibler2010peregrine} suggested and confirmed that MI can be also triggered from a single sideband only, i.e., from asymmetrical conditions. We will adopt this approach in this work to numerically and experimentally trigger the MI for surface gravity waves in an asymmetric manner, using AB-type and Stokes waves, and analyze the subsequent development of the unsteady wave train. 

\section{Experimental and numerical preliminaries} 

The experiments have been conducted in a state-of-the-art wave flume with the dimensions 30 $\times$ 1 $\times$ 1 m$^3$. Details regarding the apparatus and wave gauges used to ensure high resolution along the wave propagating direction can be found in \cite{he_galilean-transformed_2022}. The perturbed Stokes wave analyzed has the amplitude $a=0.011$ m, wave frequency $\omega=3\pi$ s$^{-1}$, and consequently a wavenumber of $k=9.08$ m$^{-1}$ for the water depth of 0.7 m. The boundary conditions to generate modulationally unstable waves are determined either following the surface elevation profile of an exact AB to second-order in steepness, evaluated at $x^*=-15$ m and defined as 
\begin{equation}
\begin{split}
\eta_{AB}(x^*,t)=&\textnormal{Re}\left(\psi\left(x^*,t\right)\exp\left[i\left(kx^*-\omega t\right)\right]\right) \\
&+\textnormal{Re}\left(\dfrac{1}{2}k\psi^2\left(x^*,t\right)\exp\left[2i\left(kx^*-\omega t\right) \right]\right). 
\end{split}
\label{se}
\end{equation} 
In order to investigate the frequency-asymmetric AB surface evolution, we apply a high- and a low-pass frequency filter to exclude all frequnecies higher or lower than the peak frequency, which is at $f=1.5$ Hz, respectively.
The alternative classical second-order Stokes wave perturbation approach from a single-sideband perturbation only is parametrized as the following
\begin{equation}
\eta_{PS}(t)=a\cos\left[-\omega t\right]+\dfrac{1}{2}ka^2\cos\left[-2\omega t\right]+\epsilon(t),
\label{se2}
\end{equation}
where $\epsilon(t)=\tilde{a}\cos\left[-\omega\left(1\pm\delta\right)t\right]$. Note that Stokes wave parameters $\tilde{a}=0.1 a$ and $\delta=\delta_{\textnormal{max}}=ak=0.1$ have been chosen to meet the AB surface elevation model, as defined in Eq. (\ref{se}), and satisfying the maximal instability growth rate conditions. The perturbation of the Stokes waves by injecting two side-bands is trivially achieved for $\epsilon(t)=\tilde{a}\cos\left[-\omega\left(1+\delta\right)t\right]+\tilde{a}\cos\left[-\omega\left(1-\delta\right)t\right]$.

Part of the simulations considering long-term evolution dynamics advance the time-like NLSE (\ref{NLSE}) in space using the fourth-order Runge-Kutta and pseudospectral methods \cite{yang2010nonlinear, he2022experimental}, ensuring  a high numerical accuracy. The spatial step length is $ dx = 0.0125 $ m and the temporal resolution is $dt = 0.2$  s, which are selected to also maintain stability of the simulations. To mitigate periodic boundary effects, the simulation domain size is set to $300$ s, but only the central $60$ s of data are extracted for spectral analysis and plotting purposes. Numerical dependence tests confirm that the chosen numerical steps, resolution, and domain size yield to high-fidelity results for the NLSE integration. The extraction of the wave envelope is achieved after the application of the Hilbert transform to the high- or low-pass frequency filtered AB surface elevation or perturbed Stokes waves, which are both described above.

We also employ fully nonlinear hydrodynamic simulations for validation purposes. In order to capture the complete dynamics of the MI, these simulations were conducted based on the enhanced spectral boundary integral (ESBI) method. The ESBI framework accounts for high-order nonlinear effects and provides an accurate representation of nonlinear wave interactions. The numerical method used for these simulations is detailed in the previous work \cite{wang2021modeling}. Adopting the highest 7th order of convolution within the current study is particularly suited for simulating complex wave interactions \cite{wang2015numerical,wang2021modeling,wang2023enhanced}. Indeed, such fully nonlinear framework provides a robust tool for understanding the formation and evolution of extreme wave events. The  injection of the boundary condition (\ref{se2}) is achieved through an optimized pneumatic wave maker \cite{clamond2005efficient}, which introduces a prescribed varying pressure field at the water surface. Several iterations on the pressure field correction are adopted to ensure the wave generation accuracy, similarly to the physical wave maker. 

\begin{figure*}[ht]
\includegraphics[width=0.65\textwidth]{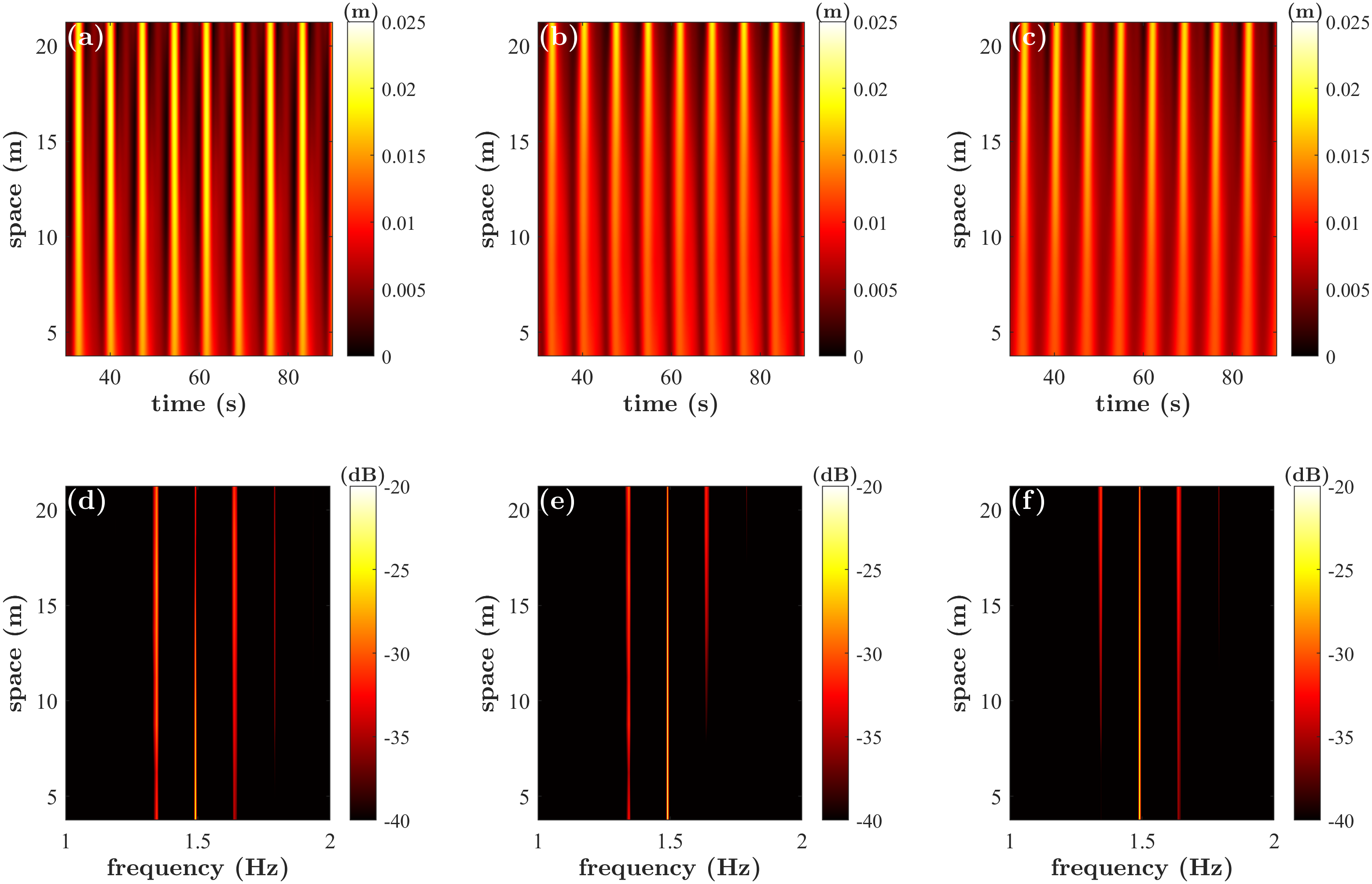}
\caption{Experimental wave evolution of (a) AB solution, (b) AB envelope including left sideband(s) only, and (c) AB envelope including right sideband(s) only. (d), (e), and (f) Corresponding power spectra of wave elevation data from (a), (b), and (c), respectively.}
\label{fig1}
\end{figure*}

\section{Results and Validation}
In this Section, we report and discuss the results of the laboratory experiments performed starting from a standard AB at its early stage of evolution and followed by the cases of high- or low-pass frequency filtering with respect to the peak frequency. After that, we proceed with the corresponding analogous case by considering the classical double and then single-sidebanded perturbation of a second-order Stokes wave. 

Fig. \ref{fig1} shows the results of wave envelope and spectral evolution for the Akhmediev breather case. We recall that the wave envelope is extracted from the water surface elevation by means of the Hilbert transform. We also align all measurements by the value of the deep-water group velocity value $c_g=\dfrac{\partial\omega}{\partial k}$. The beating of the AB wave field is clearly visible while the maximal compression occurs around 15 m from the wave maker, as suggested from theory. Note that deviations of breather experiments with respect to NLSE theory has been discussed in \cite{shemer2013peregrine,Slunyaevetal2013}.

\begin{figure*}[ht]
\includegraphics[width=0.65\textwidth]{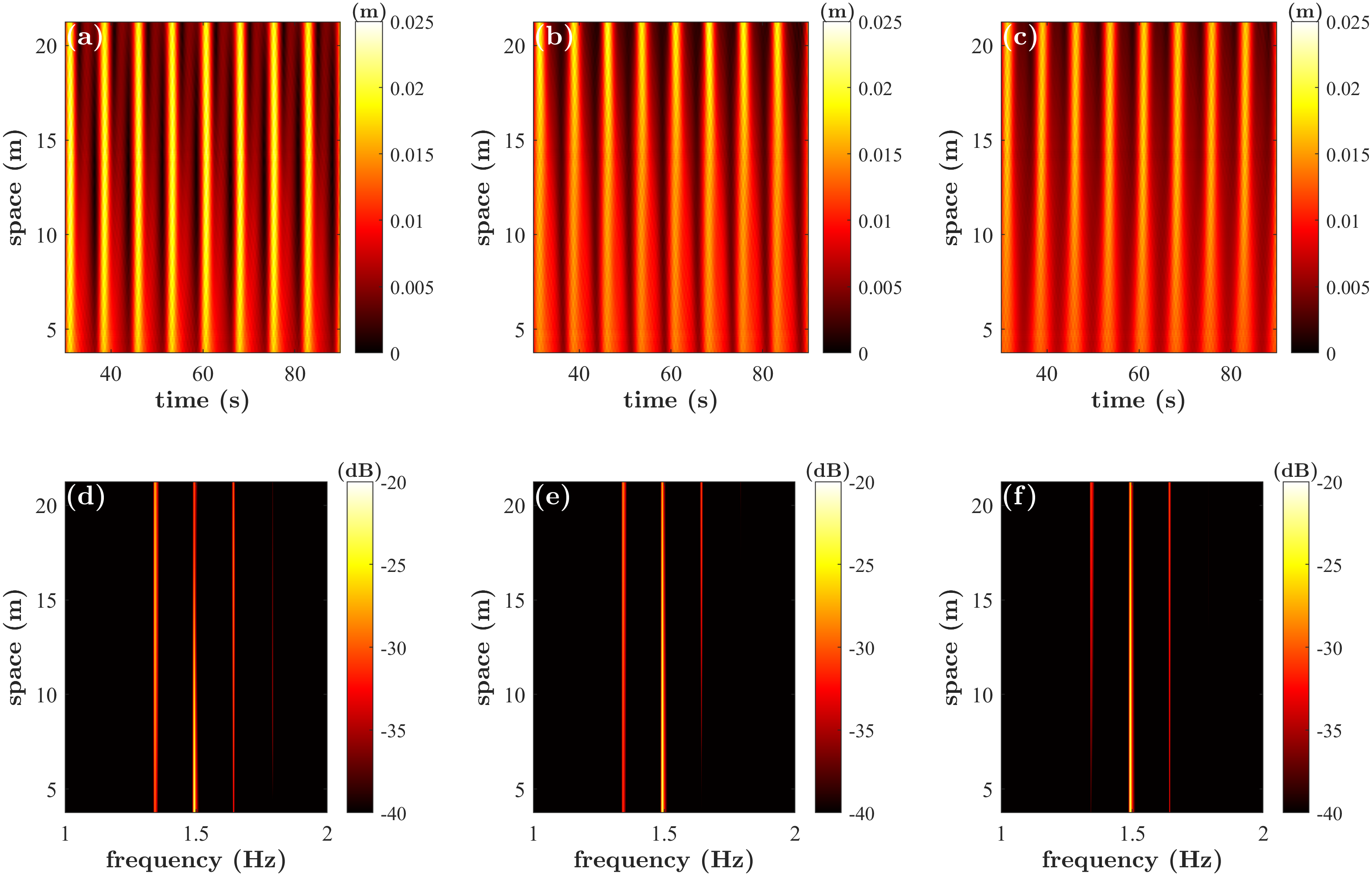}
\caption{Fully nonlinear numerical wave tank ESBI simulations corresponding to all respective cases in Figure \ref{fig1}.}
\label{fig2}
\end{figure*}

Interestingly, a focusing of the wave field can be also observed when initiating the AB dynamics with the same boundary conditions, however, ignoring either the higher (Fig. \ref{fig1} (b)) or the lower (Fig. \ref{fig1} (c)) frequencies with respect to the peak frequency, located at $f=1.5$ Hz, in the boundary conditions. The main difference is that both evolution cases of the unstable nonlinear wave fields are delayed compared to the pure AB dynamics. Note that the initial AB-type sideband cascade is not yet developed when defining the boundary conditions at $x^*=-15$ m, and thus, the contribution of the higher sideband pairs is not substantial at that early stage of evolution and as injected to the wave maker, despite being present. Also worth mentioning is that the omitted opposite sideband appears later during the wave propagation. Such retardation process as a result of non-idealized breather boundary conditions for water waves has been already discussed in \cite{chabchoub2020phase,he_extreme_2022}. The experimental results are validated against fully nonlinear numerical wave tank ESBI simulations and as depicted in Fig. \ref{fig2}. 

Indeed, the ESBI-based numerical wave tank surface elevation simulations accurately capture the nonlinear dynamics of the AB envelope solution together with the other two variants, demonstrating excellent agreement with the experimental data. This can be also noticed in all wave fields' spectral dynamics. 

\begin{figure*}[ht]
\includegraphics[width=0.65\textwidth]{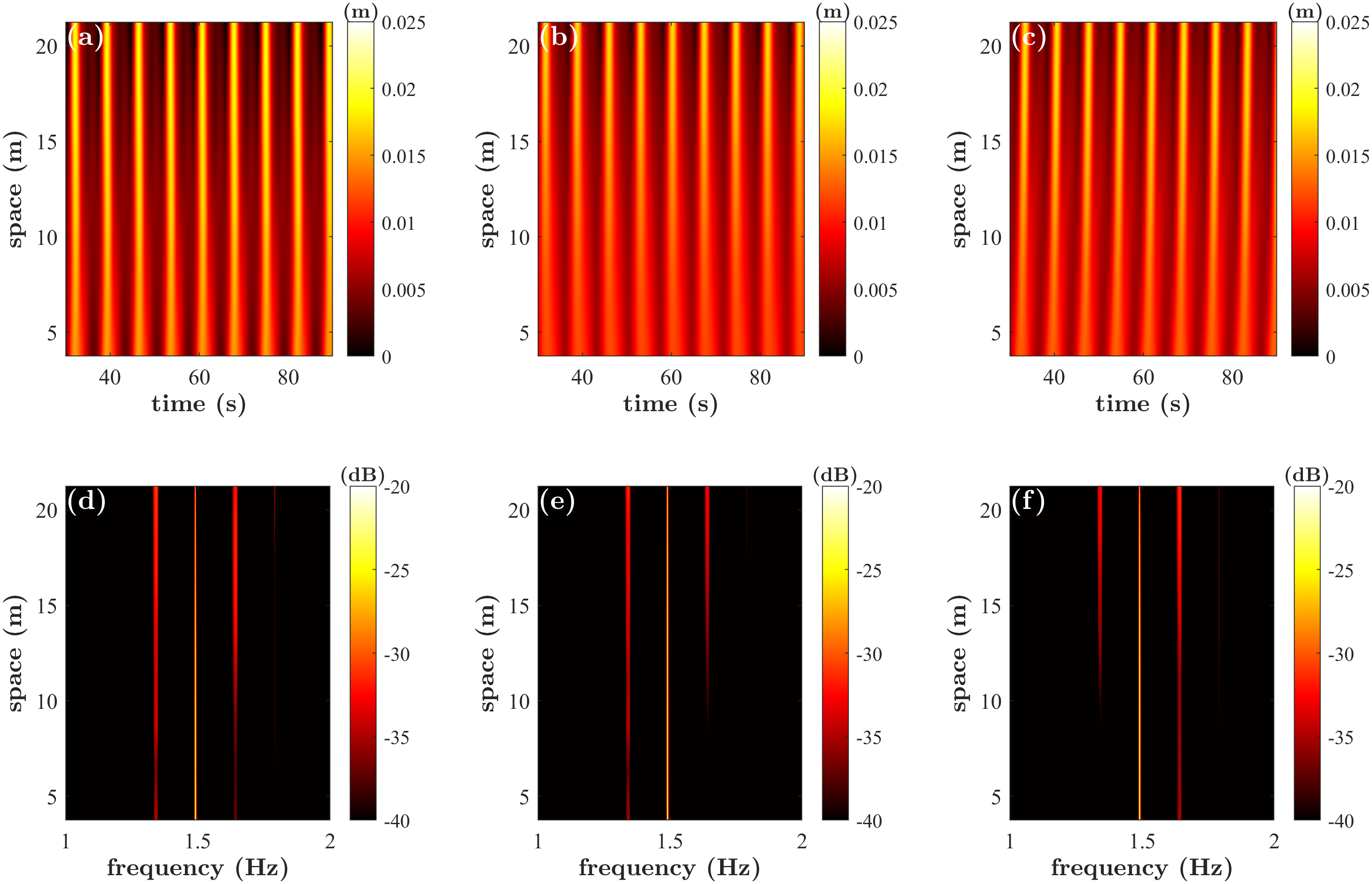}
\caption{Experimental wave envelope evolution of (a) Stokes wave perturbed with a symmetrical side-pair, (b) Left sideband perturbation only, and (c) Right sideband only. (d), (e), and (f) correspond to the power spectra of wave elevation data from (a), (b), (c), respectively.}
\label{fig3}
\end{figure*} 

As next, we perform analogical experiments, however, by considering a perturbed second-order Stokes wave, see (\ref{se2}). Fig. \ref{fig3} shows the results. 

\begin{figure*}[ht]
\includegraphics[width=0.65\textwidth]{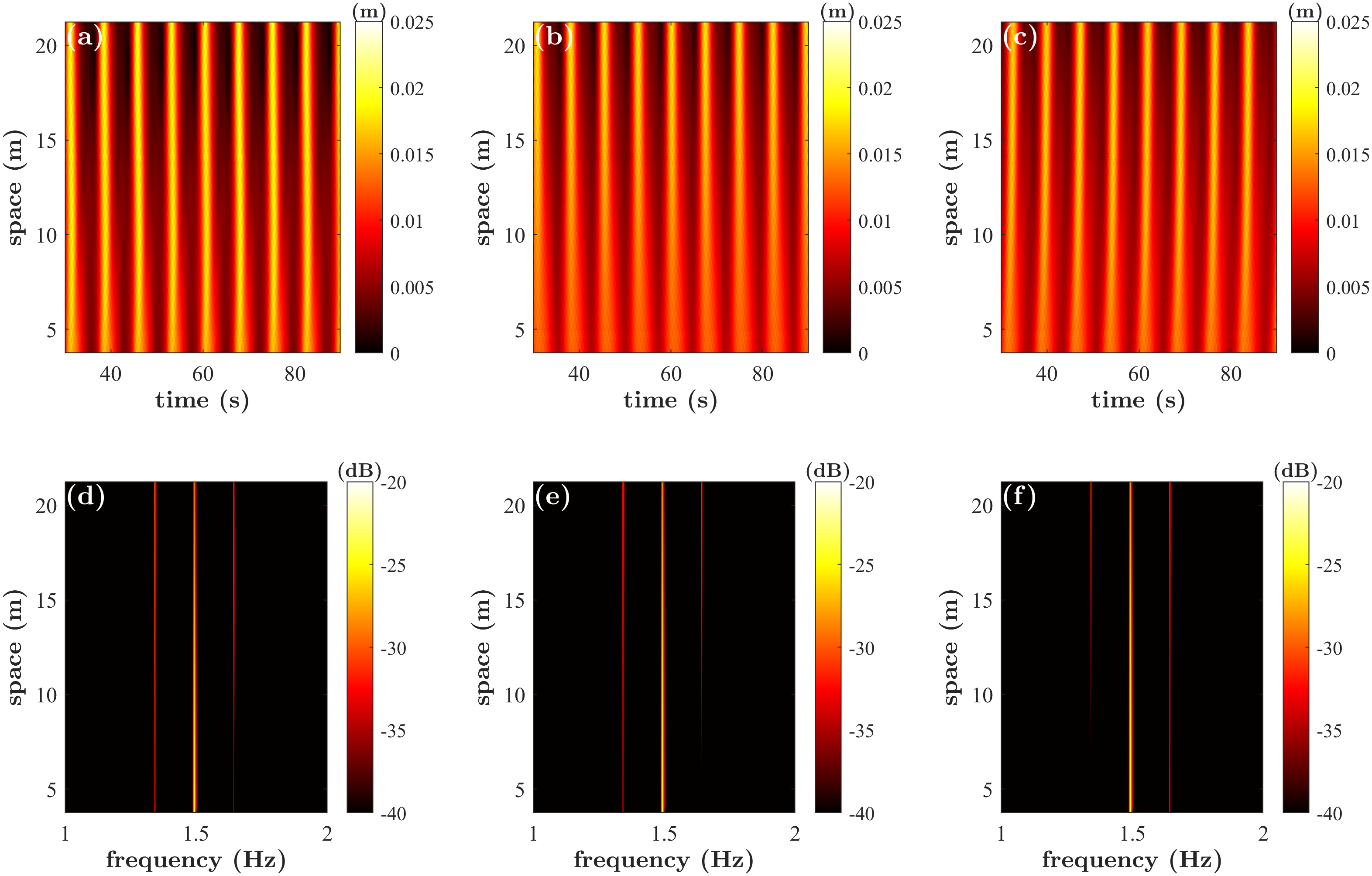}
\caption{Fully nonlinear ESBI simulations corresponding to Figure \ref{fig3}.}
\label{fig4}
\end{figure*}

Also in this case, we can distinctly observe the same wave focusing evolution trend, similarly to the AB case, that us also when injecting one sideband perturbation only. This produces the  same retardation of the focusing process as observed in the AB-type experiments. The same also applies to the ESBI simulations shown in Fig. \ref{fig4}. 

This underlines not only the accuracy of our numerical wave tank simulations, but also and once again, the sufficiency of an asymmetric one sideband only perturbation to trigger a MI-type nonlinear focusing of a quasi-regular wave field.

\begin{figure*}[ht]
\includegraphics[width=0.65\textwidth]{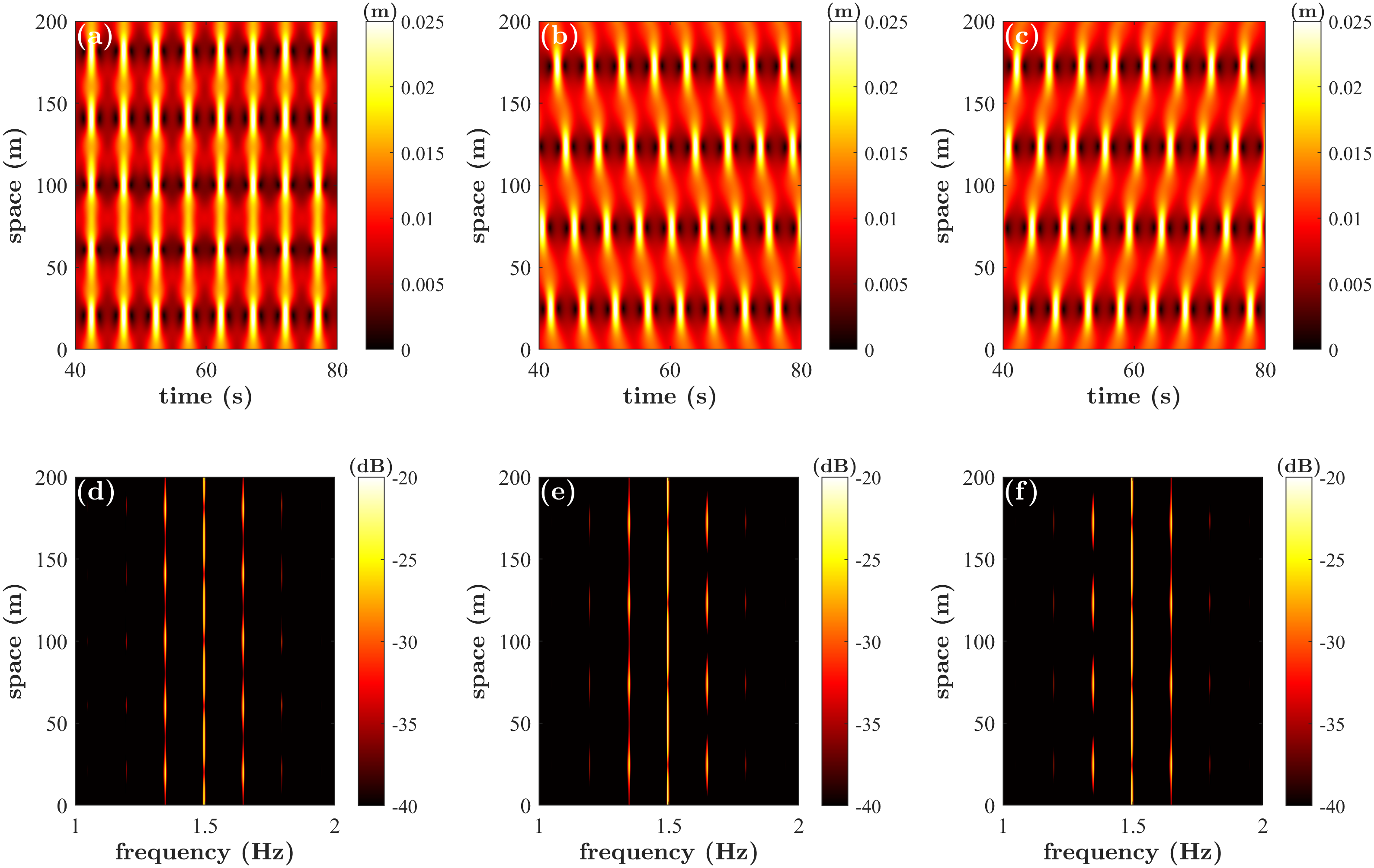}
\caption{Long-running numerical NLSE simulations corresponding to the cases described in Fig. \ref{fig3}.}
\label{fig7}
\end{figure*}

We now turn our attention to the long-term evolution of MI process, which implies a recurrent focusing process related to the FPUT recurrence. The latter process has been experimentally studied in hydrodynamics \cite{tulin1999laboratory,kimmoun2016modulation,trillo2016experimental,kimmoun2017nonconservative,eeltink2020separatrix}. In this connection, we only study the case of unstable second-order Stokes waves. The NLSE simulations results of the three types of initial perturbations, described in Fig. \ref{fig3} are depicted in Fig. \ref{fig7}. 

In addition to the slight delay in the focusing between the double-sidebanded and single-sidebanded excitations, already observed in the experiments, we can also notice a particular shift in the recurrent cycles, which appears to be due to the slight change of the value of group velocity in the asymmetric excitation. We remind that all measurements have been aligned by the value of the group velocity $c_g$. Moreover, there is almost no distinction in recurrence period between the respective single-left and single-right sideband perturbation cases. 
 
\begin{figure*}[ht]
\includegraphics[width=0.65\textwidth]{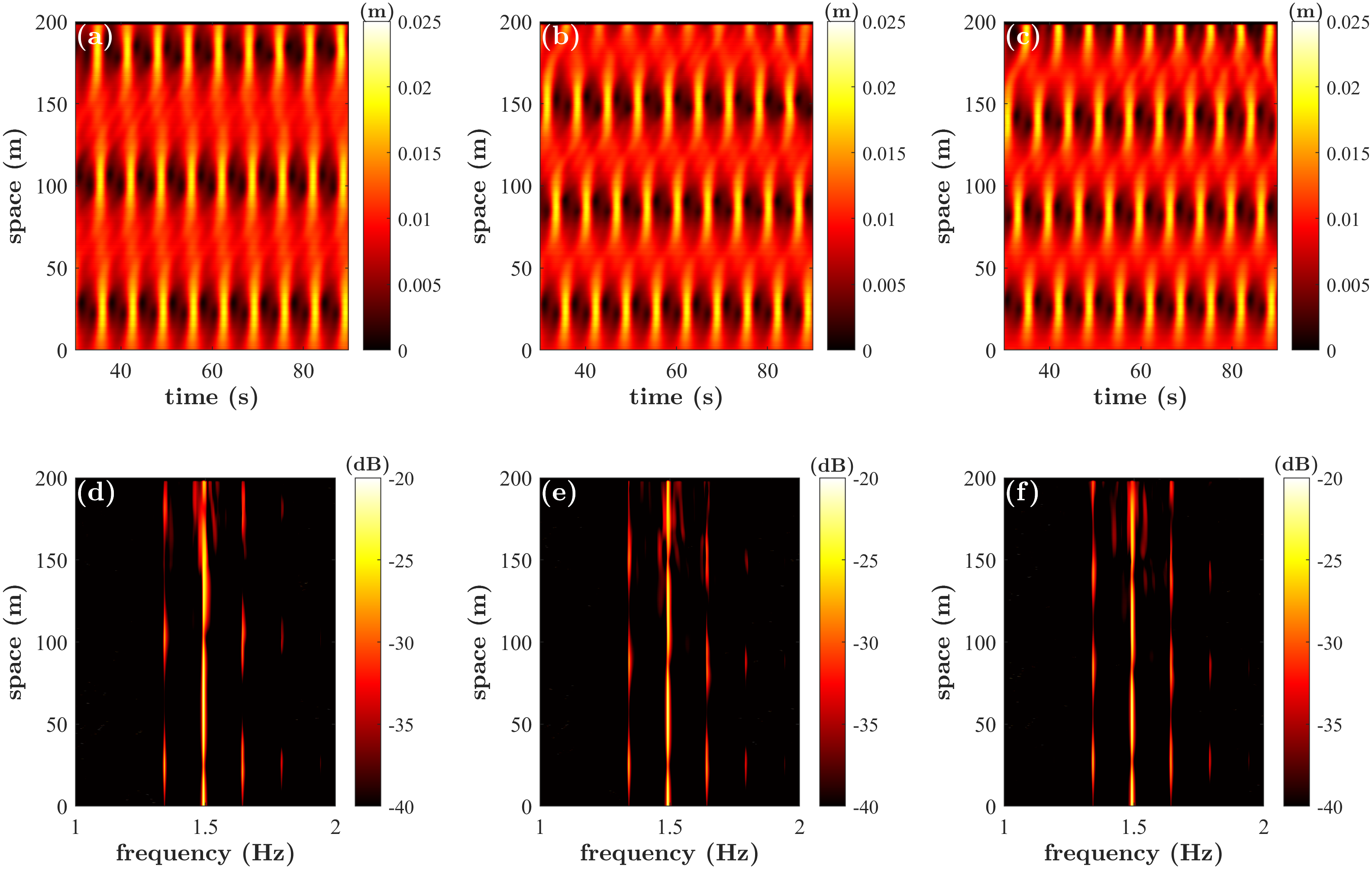}
\caption{Long-running fully nonlinear ESBI simulations corresponding to Fig. \ref{fig3} and \ref{fig7}.}
\label{fig8}
\end{figure*}

The same features are also observed in our fully nonlinear simulations pictured in Fig. \ref{fig8}. 

However, we can notice that the first focusing cycle is clearly further lagged compared to the NLSE predictions and so also the recurrent focusing period, which is clearly longer. This is not surprising and in agreement with previous benchmark studies involving the NLSE and higher-order frameworks \cite{Slunyaevetal2013,shemer2013peregrine,gomel2023mean}. 

In addition and differently to the NLSE simulations, we can distinguish a difference in wave focusing lifetime of the recurrence when comparing the the initial left to the right sideband perturbation of the Stokes wave. In fact, the retardation in the evolution and the recurrence between the right-sidebanded compared to the left-sidebanded perturbed case is clearly distinguishable. We conjecture that this is the result of the meanflow contribution, which is not considered in the symmetric NLSE framework and known to favour the growth of the left sideband. This results into an asymmetry in the shape of the focused wave groups \cite{tulin1999laboratory,Slunyaevetal2013,shemer2013peregrine,eeltink2020separatrix,gomel2023mean}. We would also like to highlight that the same characteristics in the recurrence-shift and trends are also observed for the spectrally truncated ABs, but not discussed in the paper for brevity.

\section{Conclusion}
Our experimental investigation shows that a hydrodynamic MI process can be triggered either from a one single sideband perturbation only, i.e., from a two-wave system or a spectrally truncated  AB through the application of a high- or low-pass frequency filtering from the peak frequency, and at an early stage of wave group compression. The experiments are in excellent agreement with the fully nonlinear hydrodynamic ESBI-based numerical wave tank simulations.´We emphasize that NLSE simulations predict the same experimental wave evolution trends. However, due to the higher accuracy of the ESBI framework with the respect to the order of nonlinearity, we decided not to add these in the manuscript. Long-running NLSE and ESBI simulations of the perturbations processes show a distinct phase shift in the recurrent cycles during the FPUT focusing recurrence. The numerical wave tank simulations reveal a distinct delay in the right-sidebanded perturbed Stokes wave evolution, which we attribute to the deep-water meanflow contribution. Our work may motivate future experimental studies, which will also incorporate the effects of dissipation and forcing for similar MI-type initializations. Moreover, we believe that asymmetric instability excitations in irregular wave fields require further attention to understand its effects on quasi four-wave resonant interactions in nonlinear dispersive media.

\section*{References}

% If in two-column mode, this environment will change to single-column format so that long equations can be displayed. 
%\begin{widetext}
%$$\mbox{put long equation here}$$
%\end{widetext}

% Figures should be put into the text as floats. 
% Use the graphics or graphicx packages (distributed with LaTeX2e).
% Put the label that you will use with \ref{} command in the braces of the \label{} command.
%
% \begin{figure*}
% \end{figure*}

% \begin{table}
% \end{tabular}
% \end{table}

%\begin{acknowledgments}
% Put your acknowledgments here.
%\end{acknowledgments}


\bibliographystyle{aipnum4-1} 
%\nocite{*}
\bibliography{reference}

\end{document}